\documentclass[twocolumn,pra,a4paper]{revtex4}
\usepackage{amsmath}
\usepackage{amssymb}
\usepackage{citesort,epsfig}
\begin{document}

\title{A statistical theory of nucleation in the presence of
uncharacterised impurities}
\author{Richard P. Sear}
\affiliation{Department of Physics, University of Surrey,
Guildford, Surrey GU2 7XH, United Kingdom,
{\tt r.sear@surrey.ac.uk}}

\begin{abstract}
First order phase transitions proceed via nucleation. The
rate of nucleation varies exponentially with the free-energy
barrier to nucleation, and so is highly sensitive to variations
in this barrier. In practice, very few systems are absolutely pure,
there are typically some impurities present which are rather poorly
characterised. These interact with the nucleus, causing
the barrier to vary, and so must be taken into account.
Here the
impurity-nucleus interactions are modelled by random variables.
The rate then has the same form as the partition function
of Derrida's Random Energy Model, and as in this model
there is a regime in which the behaviour is non-self-averaging.
Non-self-averaging nucleation is nucleation with a rate that
varies significantly from one realisation of the random variables to
another. In experiment this corresponds to variation in the nucleation rate
from one sample to another. General analytic expressions are obtained for
the crossover from a self-averaging to a non-self-averaging rate
of nucleation.
\end{abstract}
\maketitle

\section{Introduction}

Nucleation has long been known to be very sensitive to impurities.
Very pure water can be cooled
to tens of degrees below freezing, 0C at atmospheric pressure,
before it crystallises, but in practice the water in our freezers
freezes at only a little below 0C \cite{debenedetti}.
The 
crystals of ice in our freezer
presumably nucleate heterogeneously, in contact with some
unknown impurity in the water. The nucleus of water may
be only a few water molecules across and so is only a nanometer
or so across. Thus, even impurities only a nanometer across
can interact with the nucleus and so greatly
reduce the free-energy barrier to nucleation. The impurity
may of course be much larger.
Often we know little
of the impurity that is
providing a surface where the nucleus of ice can form at
a much lower free-energy cost than in the bulk.
Here, we circumvent the problem that the impurities are typically
uncharacterised, by using a statistical theory. We address the
question:
Under what conditions
can chance variations from sample to sample in the impurities present,
cause the nucleation rate to vary significantly from sample to sample?
That is we develop a theory that links an observable,
the variability of nucleation rate, with the variability
of the impurities at microscopic length scales.

Given the ubiquitous nature of this problem
of heterogeneous nucleation occurring on uncharacterised impurities,
relatively little theoretical work has been done. 
Karpov and Oxtoby \cite{karpov94,karpov96} have considered
nucleation in the presence of random static disorder,
and Harrowell and Oxtoby \cite{harrowell93} looked at
the effect of the distribution
of time scales present in glasses.
But this work did not address the problem of sample to sample variability,
and little theoretical work has been done for a number of years.
Castro and coworkers \cite{castro99,castro03} studied the process
that follows nucleation, namely growth. See also
Ref.~\cite{karpov95}.
The pattern of growth
depends on whether nucleation occurs continuously throughout
the process of phase transformation or only at a few sites
near the start of the process. We find sample to sample
variability occurs when one or a few sites have unusually low
nucleation barriers and so there should be a correlation between
the pattern of growth (and hence the final distribution of grain
sizes if the new phase forming is crystalline) and sample to sample
variability in the nucleation rate. Castro and coworkers
consider only growth, they did not explicitly consider nucleation, and
they did not
consider sample to sample variability.

Just as Karpov and Oxtoby did \cite{karpov96}, we will consider nucleation
in the presence of disorder. We will model the system as a nucleus
interacting with random disorder, i.e., the free energy of the
nucleus will contain a part that is a random variable. Essentially,
faced with a situation where we know the free energy barrier
to nucleation depends on its interaction with species unknown,
we realise that it is not possible to base a theoretical
description on precise knowledge and make a plausible simple guess.
Individual interactions
are modelled by random variables with some mean and standard
deviation and the system is then characterised just by these two
numbers.

The rate of nucleation at a site is proportional to
the exponential of minus the free-energy divided by the thermal
energy $k_BT$. See the book of Debenedetti \cite{debenedetti}
or the review of Oxtoby \cite{oxtoby92} or of Kashchiev
and van Rosmalen \cite{kashchiev03}
for an introduction to nucleation. Thus the rate at a particular
site is proportional
to the Boltzmann factor of the nucleus at that site and so a sum over
different sites with different free-energy barriers
has the form of a sum over Boltzmann weights. This is of course the
form of a partition function; a partition function
of a system where the energies are random variables. Such a
system is called the Random Energy Model (REM) and was first
proposed and studied by Derrida \cite{derrida80}. He
was using it as a simple model for a glass. We can take
over much of the analysis of the REM done by Derrida and apply
it to our system. Most importantly, at low temperatures the REM
is not self-averaging: different realisations of the disorder
give rise to significantly different partition functions. In our system
the analogue of the partition function of the REM is the total
rate of nucleation, and different realisations
correspond to different samples prepared in the same conditions.
So, we have a
regime in which the rate is
not self-averaging: it differs significantly from sample to sample.
Note that this is distinct from variability in properties
such as the time until the first nucleus appears. As the crossing
of a nucleation barrier is a random process the time it takes
will always be a random variable, but if there is little or no
variability in the free-energy barrier the {\em rate} itself
will self-average and so not vary
from sample to sample.
Having recognised that our problem is isomorphic
to Derrida's REM we have a model for the experimental
observation of sample-to-sample variability. This model
allows us to obtain quantitative relations between the 
width of the distribution of the free-energy barriers
to nucleation, the number of
nucleation sites, and the sample-to-sample variability.

\begin{figure}
\caption{
\lineskip 2pt
\lineskiplimit 2pt
(Colour online)
Schematic representation of a nucleus represented
by a 3 by 3 by 3 cube of dark blue monomers, in contact
with a flat surface composed of 2 types of monomers: light
and dark yellow.
\label{model}
}
\vspace*{0.3in}
\begin{center}
\epsfig{file=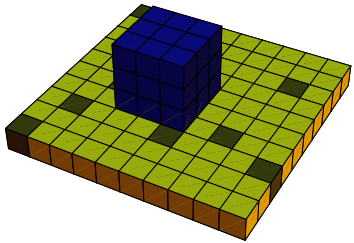,width=2.0in}
\end{center}
\end{figure}

The next section is a very general study of nucleation
with a free energy barrier that contains a term that is a random
variable.
The number of nucleation sites $N_s$ is fixed, although
our theory can be generalised to deal with varying amounts
of impurity nucleation sites, see section \ref{turnbull}.
Section \ref{surf} is devoted to the study
of an explicit model of a disordered
system: a surface composed of two types of monomers that are
distributed at random. Figure \ref{model} is a schematic
of this model. We show how this random distribution of monomers
leads to a random term in the free energy of a nucleus
in contact with the surface and obtain an explicit expression
for the width $w$ of the distribution of free-energy barriers.
The model of Fig.~\ref{model} is just one possible system that
results in a random term in the free-energy barrier to nucleation,
we can envisage many others. Indeed other activated processes
with the same exponential dependence on the height of a free-energy
barrier, such as protein unfolding \cite{searunf}, have essentially
the same behaviour in the presence of disorder. Disorder
can be a model not only for uncharacterised impurities but also
for very complex environments such as that inside a living cell.
Section \ref{secbayes} outlines the use of Bayes's theorem
to estimate the nucleation rate from a small number of
observations of nucleation. This is useful as if the nucleation
rate can be estimated for two different samples and shown
to be different in these two samples, the experimental
system must be in the non-self-averaging regime.
The last section is a conclusion.

\section{General theory}
\label{genth}

Nucleation is an activated process
\cite{debenedetti,oxtoby92,kashchiev03}. As such, its rate
has an exponential dependence on the free-energy
barrier to nucleation, $\Delta F^*$, the free-energy of
the critical nucleus. The critical nucleus is, by definition,
the nucleus at the top of the barrier to nucleation
\cite{debenedetti}. Thus, if at site $i$ of the system,
the free energy barrier is $\Delta F^*_i$, and the frequency
of attempts at unfolding is $\nu_i$, then the rate of nucleation
at the site $i$ is
\begin{equation}
R_i=\nu_i\exp\left(-\Delta F^*_i\right).
\label{ri}
\end{equation}
We will assume that the attempt frequency $\nu_i$
is only weakly dependent on $i$ and so treat it as a constant: $\nu_i=\nu$.
As $\Delta F^*_i$ is exponentiated,
if it varies appreciably then its variation dominates that of $\nu_i$
which can then be neglected. 
We use units such that the thermal energy $k_BT=1$.
If the system consists of $N_s$ possible sites for
nucleation then the average nucleation rate per site is
simply
\begin{eqnarray}
R&=&N_s^{-1}\sum_{i=1}^{N_s} R_i\\
&=&N_s^{-1}\nu\sum_{i=1}^{N_s}\exp\left(-\Delta F^*_i\right).
\label{hetnuc}
\end{eqnarray}
Thus to calculate the nucleation rate we require the $N_s$
values of the nucleation barrier at all possible nucleation sites.

Often, the system of interest is complex, or poorly characterised
with unknown impurities present. Then, we have little hope
of determining all the $N_s$ values of $\Delta F^*_i$.
To deal with these situations we resort to a statistical
approach: we guess the values of $\Delta F^*_i$. We do
this by picking the $\Delta F^*_i$ from a probability
distribution function that is characterised by two parameters,
its mean $m$ and standard deviation $w$. These two parameters
can in turn be obtained from a model, estimated from experimental
data, or simply varied to see what qualitative behaviour is possible.
We estimate them from a specific model in section \ref{surf}.

It is convenient to express the $\Delta F^*_i$ as a mean plus
a deviation,
\begin{equation}
\Delta F^*_i=m+\delta_i,
\label{dff}
\end{equation}
where $\delta_i$ is a random variable with zero mean, it is the deviation
of the nucleation barrier at site $i$ from its mean value $m$.
Taking the probability distribution of $\delta_i$, $p(\delta_i)$,
to be a Gaussian, we have
\begin{eqnarray}
p(\delta_i)&=& \frac{\exp\left[-\delta_i^2/(2w^2)\right]}
{\left(2\pi w^2\right)^{1/2}}.
\label{pdf}
\end{eqnarray}
Using Eq.~(\ref{dff}) for $\Delta F^*_i$ we can write Eq.~(\ref{hetnuc})
as
\begin{equation}
R=N_s^{-1}\nu\exp(-m)\sum_{i=1}^{N_s}\exp\left(-\delta_i\right).
\label{remnuc}
\end{equation}
Now, with the $\Delta F^*_i$ independent
random variables, the rate of Eq.~(\ref{remnuc}) is, 
except for constant factors, equivalent to
the partition function of the Random Energy Model (REM) of
Derrida \cite{derrida80}. The REM is a simple and well
understood model of glasses and other disordered systems that
undergo a transition to a state that is non-self-averaging.

Just as in the REM the average partition function can be obtained,
we can obtain the average, over realisations of the
disorder, of the nucleation rate $R$,
\begin{eqnarray}
\langle R\rangle&=&N_s^{-1}\nu\exp\left(-m\right)
\langle\sum_{i=1}^{N_s} \exp\left(-\delta_i\right)\rangle\\
&= &\nu\exp\left(-m+w^2/2\right).
\label{nuc2}
\end{eqnarray}
If the rate $R$ is self-averaging then for almost all realisations
$R$ will be close to $\langle R\rangle$ and the right-hand side
of Eq.~(\ref{nuc2}) will be a good approximation to the nucleation rate
of almost all realisations of our model of the surfaces inside a cell.
But if the rate $R$ is {\em not} self-averaging then Eq.~(\ref{nuc2})
will not be a good approximation and the rate $R$ will differ
appreciably from one realisation to another.
Nucleation in the presence of random static disorder
was considered by Karpov and Oxtoby \cite{karpov96} who obtained
results similar to that of Eq.~(\ref{nuc2}), but they only considered
self-averaging systems.

\subsection{Measures of non-self-averaging behaviour}

We will now look at how as the width of
the distribution of free-energy barriers, $w$,
increases the behaviour ceases to be self-averaging. Firstly,
we will look at how many nucleation sites contribute significant
amounts to the nucleation rate in a typical realisation.
If this number is large then as the sites are assumed independent
the rate is a sum of a large number of independent random
variables and so will be self-averaging, whereas if it is small
this will not be the case.

From Eq.~(\ref{remnuc})
we see that the rate $R$ is dominated by sites
with values of $n_i$ where
the product of the number of sites
and $\exp(-\delta_i)$, is a maximum. The number of sites
is simply proportional to the probability of Eq.~(\ref{pdf}).
The maximum of the product $p(\delta_i)\exp(-\delta_i)$ is at
a value of $\delta$,
\begin{equation}
\delta_{max}=-w^2.
\label{dmax}
\end{equation}
Now, the {\em average} number of sites around this value of $\delta_i$
is just $N_sp(\delta_{max})$, and because this average is a sum
over independent random variables (the $n_i$) the ratio
of the fluctuations to the mean scales as $[N_sp(\delta_{max})]^{-1/2}$.
Thus the fluctuations in the number of sites that contribute the
dominant amount to the rate, and hence the fluctuations in the rate
itself are small relative to the mean if and only if
$N_sp(\delta_{max})\gg 1$. From
Eqs.~(\ref{pdf}) and (\ref{dmax}) this is true whenever $2\ln N_s-w^2>0$.

Thus, the boundary between self-averaging and non-self-averaging regimes
is given by the equation
\begin{equation}
2\ln N_s-w^2=0.
\label{transit}
\end{equation}
Thus the rate is self-averaging if and only if the logarithm of the
number of possible sites for nucleation,
is larger than half the variance of the nucleation barrier. This
is the main result of this work. It is a very general result, i.e.,
it applies generally to activated processes in a random or near-random
environment. Our conclusions
here apply to any process
with a rate given by an equation of the form of
Eq.~(\ref{hetnuc}). In the next section we will give the example
of heterogeneous nucleation at a disordered
surface and in Ref.~\cite{searunf}, we showed that it held for a model
of protein unfolding {\it in vivo}.

In the non-self-averaging regime, a single unfolding site can be responsible
for a significant fraction of the entire rate.
This site must of course be the site with the lowest,
i.e., most negative,
value of $\delta_i$. We denote this lowest
value by $x$. We can easily find an estimate for $x$,
which we call $\delta_{ev}$.
It is simply the value of $\delta$ at which the mean number density,
$N_sp(\delta_i)$, of sites drops below 1. This is easy to see:
it cannot be much below the value of
$\delta$ for which $N_sp(\delta)\approx 1$
as there are rarely any sites at all
below this value and it cannot
be much above it as for these values of $\delta$ there are many sites.
Thus, we have that
$\delta_{ev}$ satisfies the equation $N_sp(\delta_{ev})=1$, and
so is given by
\begin{equation}
\delta_{ev}=-\left(2\ln N_s\right)^{1/2}w,
\label{dev}
\end{equation}
where to obtain this result we ignored the denominator of Eq.~(\ref{pdf}).

So when a single site dominates the rate $R$,
and has a value of $\delta_i$ close to $\delta_{ev}$,
the rate is approximately
\begin{equation}
R_T\simeq N_s^{-1}\nu\exp\left(-m+
\left(2\ln N_s\right)^{1/2}w\right),
\label{rtyp}
\end{equation}
using Eq.~(\ref{dev}) in Eq.~(\ref{remnuc}).
Note that $R_T\ll \langle R\rangle$
for large widths; $\langle R\rangle$ increases as the exponential
of $w^2$, Eq.~(\ref{nuc2}), whereas $R_T$ increases as only
the exponential of $w$.
Equation (\ref{dmax}) tells us that at, for example $w=6$
the maximum contribution to the average rate, $\langle R\rangle$,
comes from sites
with values of $\delta$ around $\delta_{max}=36$.
At these values of $\delta$
the probability density, Eq.~(\ref{pdf}) is close to $10^{-9}$.
Thus even for $N_s=10^8$ there is on average less than one site
at values of $\delta$ close to $\delta_{max}$.
For $N_s=10^8$ most realisations have no sites
around $\delta_{max}=36$, and so
have values of $R$ rather less than its mean value  $\langle R\rangle$,
and closer to $R_T$. The large value of $\langle R\rangle$ is due
to a few realisations with very large values of $R$.

Our analysis started with Eq.~(\ref{ri}), the standard
expression for the rate of a barrier-crossing process. This
is only valid if there is a barrier to cross, i.e., if
$m+\delta_i$ is at least a few $k_BT$. If there are sites
present for which $m+\delta_i$ is close to zero, which
is true if $m-(2\ln N_s)^{1/2}w\lesssim 0$ (Eq.~(\ref{dev})),
then the nucleation rate at these sites will be essentially
$\nu$. In this case we would expect
these sites to dominate the nucleation rate as nuclei form
effectively immediately at these sites. The rate will then
be self-averaging if and only if the average number of these sites
in a sample is much larger than one. In the remainder of the
manuscript we will assume that $m-(2\ln N_s)^{1/2}w$ is
at least a few $k_BT$.

Also, Eq.~(\ref{rtyp}) is for the rate when it is dominated
by a single site. We would expect that often when nucleation
has occurred at a site the growing domain of the nucleated
phase will prevent the formation of further nuclei at this site.
If this is so then once the first nucleus has formed then the rate
$R$ will decrease as then only the other sites with higher
free-energy barriers to nucleation will remain. Thus
associated with non-self-averaging nucleation rates we expect
rates that are time dependent. When the rate $R$ contains
contributions from many sites, clearly the rate will only
decrease after many nuclei have formed and so any time
dependence will be much less noticeable.
The rates $R$ considered here
are therefore {\it initial} rates. As determining the time
dependence of rates requires study of the behaviour of nuclei
after they have crossed the barrier we do not consider this
time dependence here, although see Refs.~\cite{castro99,castro03}
for post-nucleation growth in systems with distributions
of nucleation barriers.

We will now perform a quantitative analysis
of the fraction of the rate due to
the site with the lowest free-energy barrier, i.e., due to the one
with $\delta_i=x$. We calculate
the average, $f_{ev}$, of the fraction of the rate due to
the site with the lowest free-energy barrier. This can
be calculated from the probability distribution function, $p_{ev}(x)$,
using
\begin{equation}
f_{ev}=\frac{\nu\exp\left(-m\right)}{N_s\langle R\rangle}
\int p_{ev}(x)\exp\left(- x\right){\rm d}x.
\label{fev1}
\end{equation}
We can simplify Eq.~(\ref{fev1}) by introducing the reduced
variable $y=x/w$. Then, from Eq.~(\ref{fev1}) and using
Eq. (\ref{nuc2}) for $\langle R\rangle$,
we obtain
\begin{equation}
f_{ev}=N_s^{-1}\exp\left(-w^2/2\right)
\int {\rm d}yp_{ev}(y)\exp\left(- wy\right),
\label{fev2}
\end{equation}
where $p_{ev}(y)$ is the probability distribution function for
the minimum value of a set of $N_s$ values taken from a Gaussian
of zero mean and unit standard deviation.
Note that although the absolute value of the rate $R$ and
of the contribution of the extreme value both depend on the mean $m$,
$f_{ev}$ does not. It depends
only on $w$, and $N_s$.

The determination of $p_{ev}(y)$ is a standard problem in extreme-value
statistics \cite{sornette}.
We start from the fact that
the probability that the minimum of $N_s$ values is $y$
is the probability that 1 of the $N_s$ sites has
a value $y$, and all
the remaining $N_s-1$ sites have larger values,
multiplied by $N_s$, as
any one of the $N_s$ sites can have the lowest value. Thus,
\begin{equation}
p_{ev}(y)=N_sp(y)p_{>}^{N_s-1}(y),
\label{pev1}
\end{equation}
where $p(y)$ is a normalised Gaussian of zero mean and unit
standard deviation, and
$p_{>}(y)$ ($p_{<}(y)$) is the probability of obtaining a number
larger (lower) than
$y$ from a Gaussian of zero mean and unit standard deviation.
We are interested in the region where $x$ is several standard
deviations below the mean, $y\ll -1$.
Now, $p_{>}=1-p_{<}$, and so as
for $y\ll -1$, $p_{<}\ll 1$, we can rewrite
Eq.~(\ref{pev1}) as
\begin{equation}
p_{ev}(y)\simeq N_sp(y)\exp\left[-N_sp_{<}(y)\right],
\label{pev2}
\end{equation}
where we replaced $N_s-1$ by $N_s$.
Also, $p_{<}(y)=(1/2)\mbox{erfc}(-y/2^{1/2})$, which
for $y\ll -1$ simplifies to
\begin{equation}
p_{<}(y)\simeq \exp\left(-y^2\right)/[
(2\pi)^{1/2}(-y)].
\end{equation}

\begin{figure}
\caption{
\lineskip 2pt
\lineskiplimit 2pt
The mean fraction, $f_{ev}$, of the rate $R$ that is due to the site
with the lowest $\delta_i$,
as a function
of the width of the Gaussian, $w$.
The solid, dashed and dotted curves are
for $N_s=10^4$, $10^8$ and $10^{12}$ sites, respectively.
\label{figfev}
}
\vspace*{0.3in}
\begin{center}
\epsfig{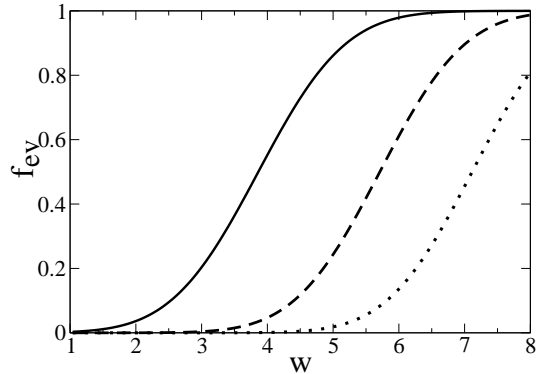}
\end{center}
\end{figure}

In Fig.~\ref{figfev} we have plotted the fraction of the rate due
to the site with the lowest
barrier, $f_{ev}$, as a function of $ w$.
We took $N_s=10^4$, $10^8$ and $10^{12}$.
For protein crystallisation \cite{durbin96}
distinct sites should be at least 1nm
apart. Then $N_s=10^8$ sites corresponds to a surface
of order 100$\mu$m$^2$. The dependence on $N_s$ is logarithmic,
varying $N_s$ by orders of magnitude does not have a marked effect.
$\ln N_s$ should nearly always be of order 10.
We see that as $w$ increases, so does $f_{ev}$.
For $N_s=10^8$, Eq.~(\ref{transit}) is satisfied for
$w=6.07$. For $w$ around this value the site
with the largest interaction energy already contributes a large
amount to the total rate, on average. This large contribution
will vary significantly from one realisation to the next,
and so the fraction of the rate due to the site with the lowest value
of the nucleation barrier
will vary substantially from realisation to realisation at large $w$.
For some realisations it will be rather larger than $f_{ev}$ and
for others it will be much smaller.
Whereas of course if $w$ is small the rate $R$ has significant
contributions from many unfolding sites and so varies weakly
from realisation to realisation,
essentially due to variations in the rate being
averaged out in accordance with the central-limit theorem.

\subsection{Variable $N_s$}
\label{turnbull}

There is data on the effect of impurities from the work of
Turnbull \cite{turnbull50} and coworkers, and that
of Perpezko \cite{perpezko84}
coworkers on nucleation from
dispersions of liquid droplets \cite{debenedetti,oxtoby92}.
These experiments were motivated by the idea that if sufficiently
small droplets could be formed some droplets
would be free of all impurities and
in those droplets the nucleation would then be homogeneous.
It is not clear that this objective was achieved
\cite{debenedetti,oxtoby92,perpezko84}.
Perpezko \cite{perpezko84} assumed that
the impurities are randomly distributed, and then the 
number of impurity particles in a droplet is given by a Poisson
distribution function. He addressed the question of random
variation in
the number of impurity particles but not that of variation in the interaction
of the impurity with the nucleus. Thus in a sense it is complementary
to this work. If we make the number of sites $N_s$ itself a random variable
but set $w=0$ then we obtain the model of Perpezko \cite{perpezko84}.
Thus if we allow the number of nucleation sites $N_s$ to be a random
variable while maintaining $w$ non-zero we have a model that can
describe both variation in both the number of impurity particles
and disorder in the surface of these particles. We leave such
a generalisation to future work.



\section{Disordered surfaces}
\label{surf}

In the previous section we merely assumed that the presence
of disorder introduced a random part $\delta_i$ into the nucleation barrier
at site $i$, and that the $\delta_i$ are drawn from a Gaussian distribution.
In this section we will start from a simple model of a disordered
surface and show that in a certain limit, a Gaussian distribution
of free-energy barriers is obtained, and obtain expressions
for the mean $m$ and width $w$, of this Gaussian,
in terms of the parameters that
characterise the surface.

Surfaces, for example of impurities,
can provide sites for nucleation. We consider a simple
planar surface formed of a plane of sites of a cubic lattice
all occupied by fixed monomers. The nucleus is taken to be a block
of monomers of single type which may be the same type as some of
those of the surface or different.
We assume that not more than one monomer can occupy a site, thus
the nucleus can be in contact with the surface and so interact
with it but it cannot penetrate the surface.
Apart from this excluded-volume interaction, the only interactions
are those between monomers in contact.
If the surface were uniform, i.e., composed exclusively
of one type of monomer then the free energy barrier to nucleation
would of course be the same at every point on the surface.
However, if the surface is composed of 2 types of monomers
that are not uniformly distributed then the barrier will vary
from point to point, depending on the numbers of monomers
of the different types that the nucleus is in contact with at
a particular point. A schematic of a cubic nucleus
in contact with such a surface is shown in Fig.~\ref{model}.

Let us call the 2 types of monomer A and B, and assume they
are distributed at random. Let monomers of type $A$ and $B$
interact with the nucleus with energies $\epsilon_A$ and
$\epsilon_B$, respectively. Then the shift in the barrier
to nucleation when the nucleus is at a site $i$ in contact
with the surface is
\begin{equation}
\Delta F^*_i=\Delta F^*_0+n_i\epsilon_A+(n_c-n_i)\epsilon_B,
\label{dfs}
\end{equation}
where $\Delta F^*_0$ is the nucleation barrier when the
the nucleus is not in contact with the surface.
$n_c$ is the total number of sites in the nucleus that
contact the surface; as the surface is taken to be planar
this number is taken to be a constant. $n_i$ is the
number of A monomers of the surface in contact with the
nucleus when the nucleus is at site $i$.
If the monomers of the surface are either A or B at random,
then the probability of any one of the $n_c$ sites
of the surface being an A-type monomer is just the fraction
of A-type monomers, which we denote by $f_A$.
Then the probability of the nucleus being in contact
with $n_i$ A-type monomers and $n_c-n_i$ B-type monomers
is just
\begin{eqnarray}
p_A(n_i)&=&\frac{n_c!}{n_i!(n_c-n_i)!}f_A^{n_i}\left(1-f_A\right)^{n_c-n_i}
\\
&\simeq& \frac{\exp\left[-(n_i-m_s)^2/(2w_s^2)\right]}
{\left(2\pi w_s^2\right)^{1/2}},
\label{pdfsurf}
\end{eqnarray}
where the
mean value $m_s=f_An_c$, and the variance of the Gaussian
$w_s^2=n_cf_A(1-f_A)$.

Using Eqs.~(\ref{dff}), (\ref{dfs}) and (\ref{pdfsurf})
we see that the Gaussian distribution
for $n_i$ becomes a Gaussian distribution for $\delta_i$
of variance
\begin{eqnarray}
w^2&=&w_s^2\left(\epsilon_A-\epsilon_B\right)^2\nonumber\\
&=&n_cf_A\left(1-f_A\right)\left(\epsilon_A-\epsilon_B\right)^2.
\label{ws}
\end{eqnarray}
The mean value of the $\Delta F^*_i$ of Eq.~(\ref{dff})
is 
\begin{equation}
m=\Delta F^*_0+n_c(f_A\epsilon_A+(1-f_A)\epsilon_B).
\end{equation}

For the nucleation rate to be non-self-averaging we require
that $w^2$ be larger than $2\ln N_s$, Eq.~(\ref{transit}).
Unless $N_s$ is extremely large or small $2\ln N_s$ will be
of order 10.
From Eq.~(\ref{ws}) we see that if the difference in interaction
energy between the 2 types of monomer, $\epsilon_A-\epsilon_B$
is a few $k_BT$, and if we have around $n_c=10$ sites of the
surface in contact with the nucleus, then $w^2$
will be around 10 to 30, providing that $f_A$ is neither
very small nor close to unity. Thus, we predict that
heterogeneous nucleation at
disordered surfaces composed of significant fractions of
different species whose interactions with the
nucleus differ by a few $k_BT$, will often be dominated by
one or a few sites. It will therefore vary appreciably between realisations.
Experimentally, this means that the rate will differ appreciably
between nominally identical samples.

Finally, for the purposes of comparison we consider
adsorption onto the surface of individual monomers.
These monomers are of the same type as those that made up the nucleus.
For
simplicity we do so in the regime
where we have much less than a monolayer, i.e., where
the number of adsorbed monomers
$\Gamma\ll N_s$. Now, we can compare the rate $R$ with the adsorbed
amount $\Gamma$ in order to get a feel for which property is more
likely to be non-self-averaging. When $\Gamma\ll N_s$ then few pairs
of adjacent sites are occupied and so we can treat each surface
site as being independent. Then $\Gamma$ is given by
\begin{equation}
\Gamma=\sum_{i=1}^{N_s}
\frac{\exp\left[\mu+n_i\epsilon_A+(1-n_i)\epsilon_B\right]}
{1+\exp\left[\mu+n_i\epsilon_A+(1-n_i)\epsilon_B\right]},
\label{gamma}
\end{equation}
where $n_i=1$ if the monomer at site $i$ on the surface is an A-type
monomer and $n_i=0$ if the monomer is a B-type monomer.
$\mu$ is the chemical potential of the monomers (in units of $k_BT$).
The variation of $\Gamma$ from realisation to realisation will
depend on $\epsilon_A$, $\epsilon_B$, $f_A$ and $\mu$.

However, this variability simply comes from the fact that
the terms in the sum of Eq.~(\ref{gamma}) take one of two values
depending on whether the monomer is type A or type B.
These two values are bounded by zero and one.
Thus we can easily obtain an upper bound for this variation 
in $\Gamma$ by assuming
the terms in the sum for $\Gamma$, Eq.~(\ref{gamma}) are either
zero or one. This corresponds to, say,
the A-type monomers always having a monomer adsorbed onto them
while the B-type monomers never have an adsorbed monomer.
For definiteness we assume that A-type monomers are the ones
with adsorbed monomers.
This approximation
will clearly overestimate the variability in $\Gamma$ but even
within this approximation the variance of $\Gamma$ is just
$f_A(1-f_A)N_{s}$ for large $N_s$. The ratio of the
standard deviation to the mean, $f_AN_s$, is then given by
\begin{equation}
\frac{\mbox{std. dev.}}{\mbox{mean}}=
\left(\frac{1-f_A}{f_A}\right)^{1/2}N_s^{-1/2}
\end{equation}
and so is small for large $N_s$ and $f_A=O(0.1)$. At least
when the adsorption is small $\Gamma$ is self-averaging.
So disorder large enough to cause the rate $R$ to
be non-self-averaging may leave other properties, e.g., $\Gamma$,
still self-averaging. As the nucleus is large, $n_c=O(10)$,
the variance in the free-energy
barrier at a site is large (it is multiplied by $n_c$ in Eq.~(\ref{ws}))
and the rate $R$ is then proportional to the exponential of this
large quantity. Both the factor of $n_c$
and the exponentiation strongly enhance the effect
of disorder and make the nucleation rate one of the most likely
properties of a system to be non-self-averaging.

\section{Determining the nucleation rate using Bayesian inference}
\label{secbayes}

In this section we will discuss the use of Bayesian inference
to determine the probable nucleation rate from measurements
of nucleation, and hence determine
whether or not two (or more) different samples have the
same or different nucleation rates. This is required as
observing the effects of disorder on nucleation is hampered by the
fact that nucleation is inherently a random process.
There is more than one way to study nucleation and inference
should be applicable to all of them, but for definiteness
and because our nucleation rates $R$ are initial nucleation
rates we study determining the rate of nucleation from
the time until the first nucleus appears.
Fortunately, the inference problem we need to solve is the same
as that given and solved as an example in chapter 3 of the textbook
of MacKay \cite{mackay}. We shall therefore give only
a brief presentation, referring the reader for details to
Ref.~\cite{mackay}.

Nucleation is due
to a fluctuation and so is random even in a completely uniform
pure system.
The time $t$ at which the first nucleus
appears is a random variable.
The probability distribution function for $t$
is an exponential,
\begin{equation}
p(t)=RN_s\exp\left(-RN_st\right).
\label{pt1}
\end{equation}
Experiments can also involve counting the number of events,
and if these events are independent this number is given by a Poisson
distribution function. For example
Galkin and Vekilov \cite{galkin99,vekilov03}
count the number of protein crystals formed. The analysis here can
also be applied to determine whether or not two Poisson distributions
have different means. If they have then that too indicates
a varying nucleation rate.

Let us consider the situation where we have two samples
that have been prepared in the same way.
If we can determine that they have different nucleation rates
then clearly we must be in the non-self-averaging regime whereas
if we examine a number of samples and they all have
indistinguishable rates then we are in the self-averaging regime.
A given sample will have some unknown total nucleation rate $RN_s$. If
we determine the time $t$ at which a nucleus appears $N_A$ times,
then we will have $N_A$ values, $t_1$
to $t_{N_A}$, drawn from the distribution
of Eq.~(\ref{pt1}). We denote this set of times
by $\{t\}$.

We now need Bayes's theorem, which is \cite{mackay}
\begin{equation}
P\left(RN_s|\{t\}\right)=\frac{p_0\left(RN_s\right)
p\left(\{t\}|RN_s\right)}
{\int p_0\left(RN_s\right)
p\left(\{t\}|RN_s\right){\rm d}\left(RN_s\right)},
\label{bayesth}
\end{equation}
where $P\left(RN_s|\{t\}\right)$ is the probability we want:
it is the probability that the rate is $RN_s$ given the set of measured
nucleation times $\{t\}$. Also, $p_0\left(RN_s\right)$
is the prior probability distribution, the probability
distribution before we made the measurements, and
$p\left(\{t\}|RN_s\right)$ is the probability of
observing the set of nucleation times $\{t\}$ given that the
nucleation rate is $RN_s$. This last probability is easily
obtained from Eq.~(\ref{pt1}) which gives the probability
of observing a single value of $t$ given the rate. As the
measurements are independent, $p\left(\{t\}|RN_s\right)$
is simply given by
\begin{eqnarray}
p\left(\{t\}|RN_s\right)&\propto &
\left(RN_s\right)^{N_A}
\Pi_{i=1}^{N_A}\exp\left(-RN_s t_i\right)\\
&\propto & \left(RN_s\right)^{N_A}\exp\left(-RN_st_s\right),
\label{pts}
\end{eqnarray}
where $t_s$ is the sum of the $N_A$ measurements
\begin{equation}
t_s=\sum_{i=1}^{N_A}t_i.
\end{equation}
The sign $\propto$ indicates that we have dropped a normalisation
constant. We can restore normalisation at the end of the calculation.

Using Eq.~(\ref{pts}) in Eq.~(\ref{bayesth}) we obtain
the probability distribution function of the rate
\begin{equation}
P(RN_s|\{t\})=cp_0\left(RN_s\right)\left(RN_s\right)^{N_A}
\exp\left(-RN_s t_s\right),
\end{equation}
where $c$ is just a constant of normalisation,
\begin{equation}
c^{-1}=\int p_0\left(RN_s\right)\left(RN_s\right)^{N_A}
\exp\left(-RN_s t_s\right){\rm d}\left(RN_s\right)
\end{equation}

We have considered a pair of randomly generated systems.
Each has $N_s=10^4$ sites with free-energy barriers
taken from a distribution with mean
$m=20$ and standard deviation $w=3$.
We generate two realisations, the first has a total
nucleation rate $RN_s=3.623\times 10^{-3}\nu$
and the second has $RN_s=1.575\times 10^{-3}\nu$.
To employ Bayesian inference we require a prior
distribution for the total rate,
$p_0(RN_s)$.
We pick a top hat function,
\begin{equation}
p_0\left(RN_s\right)=\left\{
\begin{array}{cc}
R_0^{-1}& RN_s\le R_0\\
0& RN_s>R_0
\end{array}\right. .
\label{p0}
\end{equation}
Other reasonable priors give similar results, as they should.

\begin{figure}
\caption{
\lineskip 2pt
\lineskiplimit 2pt
The probability distribution function $P$ for the total
nucleation rate $RN_s$. It is obtained
using Bayes's theorem applied to
$N_A=20$ measurements of the time $t$ at which
the first nucleus appears.
The true nucleation rates are
$RN_s=3.623\times 10^{-3}\nu$ (solid curve)
and $RN_s=1.575\times 10^{-3}\nu$ (dashed curve).
\label{bayes}
}
\vspace*{0.3in}
\begin{center}
\epsfig{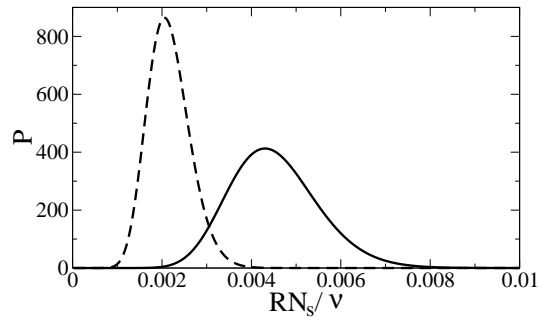}
\end{center}
\end{figure}

We have numerically generated sets of $N_A=20$ nucleation times for
both systems and used both sets of values in Eq.~(\ref{bayes}).
The two resulting probability distribution functions,
$P(RN_s|\{t\})$, are plotted in Fig.~\ref{bayes}.
We used a prior of width $R_0=5\times 10^{-2}\nu$.
Even with such a broad prior, 20 measurements are clearly
enough to demonstrate that it is very likely
that the two systems have different
nucleation rates. Thus, the use of Bayes's theorem in this way
is an effective way of determining that the rate is varying
from sample to sample, and so the rate is not self-averaging.

\section{Conclusion}

Nucleation often occurs with the nucleus interacting with,
and with a free energy strongly reduced by, impurities.
This is called heterogeneous nucleation.
Here, we have addressed the question: Under what conditions
can chance variations from sample to sample
in the impurities present,
cause the nucleation rate to vary significantly from sample to sample?
In the previous section we showed
how Bayes's theorem allows an efficient estimation of the nucleation
rate in a sample and so allows variations in this rate to be detected.
As the impurities are typically uncharacterised
and uncontrolled we resorted
to a statistical theory to model chance, i.e., random, variations in the
impurities. The impurities were modelled by quenched disorder
and we showed that the rate of nucleation has the same form as the
partition function of Derrida's Random Energy Model \cite{derrida80}.
There is a regime where the nucleation rate in
different samples prepared in the same way may be different,
where it is non-self-averaging. This occurs when the
width $w$ of the distribution of nucleation barriers is large.
The crossover
from this regime to the regime where the nucleation rate is
very similar in different samples
occurs at a width $w$ given by Eq.~(\ref{transit}).
The nucleation rate is very sensitive to disorder
in the sense that it may be non-self-averaging even when other
properties may still be
self-averaging. This is in accord with experiment where nucleation
is known to be highly sensitive to impurities \cite{perpezko84}.
Our study of a specific model of nucleation at a disordered
surface (section \ref{surf}) showed that, at least within this model,
the origin of this sensitivity lies in the fact that
the nucleus is quite large, it consists of not one but many
molecules, and that the rate is proportional to the exponential
of the free-energy barrier.
Nucleation is important in a number of fields, for example,
it is crucial for protein crystallisation \cite{durbin96}.
The crystal phase of
proteins is required for X-ray determination of their structure.

The method of direct observations of nucleation and
applying Bayes's theorem,
is not the only way of estimating the effect of disorder on
nucleation.
An alternative way
is to follow the fraction of the system that has undergone the
phase transition as a function of time. 
The evolution over time $\tau$ of this fraction,
which we denote by $X(\tau)$, 
is often described using the Kolmogorov-Johnson-Mehl-Avrami (KJMA) theory
\cite{castro99,castro03},
according to which
\begin{equation}
X(\tau) = 1-\exp(-A\tau^m),
\label{avrami}
\end{equation}
where $A$
is a constant that depends on both the rate of nucleation and the rate
of growth of the droplets/crystallites of the new phase.
Equation (\ref{avrami})
is sometimes referred to as Avrami's law.
If the nucleation rate is uniform throughout the system,
the exponent $m=d+1$ with $d$ the dimensionality
of space.
The power of $d+1$ contains a power of $d$ due to the fact that
if the growth front of the domains of the new phase is moving
at a constant velocity $v$, then the volume of a domain scales
as $(v\tau)^d$. The additional power of time comes from the fact that
for uniform nucleation the number of domains increases
linearly with time $\tau$. However, if nucleation is not uniform
but occurs at just a few sites then nucleation may occur at early
times at these sites, and then nucleation ceases as the sites
with low free-energy barriers have been `used up'.
Then the KJMA exponent $m$ equals $d$ not $d+1$.
The nucleation rates $R$ calculated here are initial rates,
when the rate $R$ is dominated by a few sites it will decrease
as they are `used up'.
Thus, as
has been discussed by Castro and coworkers \cite{castro99,castro03},
disorder can result in deviations from a simple KJMA growth
law with exponent $m=3$. See Refs.~\cite{castro99,castro03} for
calculations showing effective exponents between 2 and 3.
We would expect that non-self-averaging systems, where nucleation
occurs predominantly at one or a few sites, should exhibit
an exponent near to $m=2$.
It should be noted that they point out that $m$ alone is a not a
particularly
discriminating and that if the new phase forming is crystalline, then
the grain size distribution provides more information.

Finally, Harrowell and Oxtoby \cite{harrowell93} have discussed
the effects of the rapidly
increasing relaxation time, essentially our $\nu^{-1}$,
and heterogeneity present in a glass. Of course, glassy
systems show non-self-averaging behaviour.
Future work
could study non-self-averaging behaviour of the nucleation rate
in glasses.

It is a pleasure to acknowledge that this work has benefited
greatly from discussions with J. Cuesta.
This work was supported by The Wellcome Trust (069242).


\end{document}